\documentclass[12pt,parskip=full]{article}
\usepackage{geometry}
\usepackage[
backend=biber,
style=nature,
citestyle=nature
]{biblatex}
\usepackage{authblk}
\usepackage{physics}
\usepackage{dsfont}
\usepackage{subcaption}
\usepackage{graphicx}
\usepackage{amsmath}
\usepackage{comment}

\newcommand\numberthis{\addtocounter{equation}{1}\tag{\theequation}}

\title{Dynamical Decoherence and Memory Effects in Green Fluorescent Proteins}

\author[1,2,3]{Adam Burgess\thanks{a.d.burgess@surrey.ac.uk}}
\author[1,2,3]{Marian Florescu\thanks{m.florescu@surrey.ac.uk}}

\affil[1]{Leverhulme Quantum Biology Doctoral Training Centre, University of Surrey, Guildford, GU2 7XH, United Kingdom}
\affil[2]{Advanced Technology Institute, University of Surrey, Guildford, GU2 7XH, United Kingdom}
\affil[3]{Department of Physics, University of Surrey, Guildford, GU2 7XH, United Kingdom}

\date{}
 
\begin{document}
 \maketitle
 \begin{abstract}
The interaction of a quantum system with its surroundings has major implications for its dynamical evolution and has the ability to significantly limit our capabilities in exploiting quantum degrees of freedom for advanced applications. Understanding the environment-mediated processes by which quantum mechanical systems decohere and lose their non-classical correlations is of utmost importance and may open avenues  yet to be explored in addressing basic physics questions as well as for developing practical quantum technology applications. In parallel, in the emerging field of Quantum Biology, there is a growing body of evidence that in certain bio-molecular complexes, quantum effects  dominate over their purely classical correspondents, especially in the case of photo-excited systems for which the surrounding dielectric environment plays an essential role in the system dynamics.  Here, we explore the dynamical decoherence of chromophore within a green fluorescent protein  coupled to a finite-temperature dielectric environment, a system of significant interest due to its anomalously long coherence lifetimes. We employ the Hierarchical Equations of Motion formalism, a  non-perturbative and non-Markovian approach, and focus our study on the degree of coherence displayed by independent green fluorescent protein chromophores and the energy transfer dynamics mediated by the dielectric relaxation of the environment.  For different system architectures, we identify several striking features in the dynamics of the chromophore induced by the dielectric relaxation of the environment, resulting in strong memory effects that extend the coherence lifetime of the system. Remarkably, the complex architecture of the green fluorescent protein, which includes a cavity-like structure around the molecular chromophore system, is ideally suited to preserving the coherences in the homo-dimer system. The system dynamics generate a dynamical correlation between the coherent energy transfer between its sub-systems and the entropy production, which can lead to transient reductions in entropy, a unique feature of the non-Markovian nature of the system-environment interaction.
\end{abstract}

\begin{refsection}
\section*{Main}\label{sec1}

The tendency of quantum mechanical systems to decohere and lose their non-classical correlations - causes a significant limitation for the effective utilization of quantum technologies~\cite{Ladd2010}. Consequently, the study of decoherence in open quantum systems has flourished over the recent decades and has become a dedicated field of research in itself ~\cite{Schlosshauer_2019}. Furthermore, since the advent of microbiology, the fields of quantum mechanics, chemistry and biology have become highly entwined ~\cite{Kattnig22,Algar2019}. In the process of evolving our understanding of biological systems down to the molecular level, it becomes inarguable that quantum effects will occur in such bio-molecular systems~\cite{QBreview}. The extent to which these effects dominate or are relevant compared to purely classical effects is still a topic of much contention in the emerging field of Quantum Biology~\cite{Revisited}, but it has been argued that they may play an essential role in in the dynamics of bio-molecular complexes~\cite{Lin20GFP,Kattnig22}, in particular for photo-excited systems where the dielectric medium relaxation plays a key role~\cite{Gilmore_2005}. Moreover, within the field of quantum biology~\cite{QBreview,HuelgaBio}, open quantum systems approaches have found great utility in unveiling the complex mechanisms through which biomolecular complexes interact with their noisy and warm environments. Such approaches include studies on protein tunnelling in DNA~\cite{Slocombe2022}, magneto-sensitive cryptochromes~\cite{cryptOQS} and excitation dynamics in photosynthetic complexes~\cite{CoherencePS}. Developing an overarching understanding of decoherence and how thermal effects influence the evolution of quantum mechanical systems will undoubtedly provide valuable insights into the fledgling field of quantum biology and the maturing field of quantum technologies.

\begin{figure}
    \centering
    \includegraphics[width=0.6\textwidth]{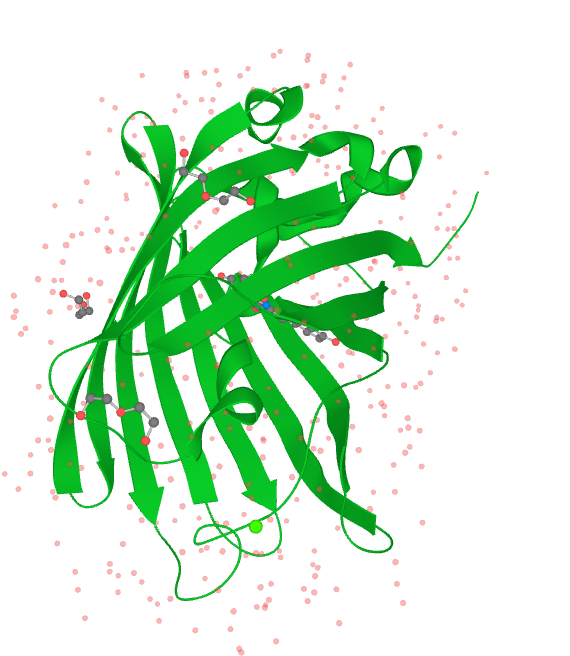}
    \caption{A computer rendering of the structure of the mEGFP~\cite{GFPOG,FPBaseLambert2019}. The $\beta$-strands around the exterior form the $\beta$-barrel structure. An $\alpha$-helix contains the chromophore that runs through the centre.  }
    \label{fig:GFPimage}
\end{figure}

A prime candidate of a bio-molecular system of significant interest in the field of quantum biology is the green fluorescent protein (GFP)~\cite{GFPOG}. The GFP has a beta-barrel structure consisting of eleven beta-strands arranged in a cylindrical configuration, with an alpha helix running through the middle that contains the covalently bound chromophore with an overall radius of 13.4\AA ~(a rendering of this structure is shown in Fig.~\ref{fig:GFPimage}). GFPs are of interest beyond purely the bio-physical interface, as they have found application in the general quantum technology field capable of entangled photon production~\cite{Shi2017}, pH sensing~\cite{GFPPh} and protein tagging~\cite{ProteinTagging}. Moreover, GFPs can be genetically modified to modulate their quantum optical properties, such as quantum yields and transition energies~\cite{Goedhart2012}. 
 Furthermore, the GFP structures can be made to dimerize, and when dimerized have an inter-chromophore distance of 27.5\AA, with less than 1\AA ~ between the GFP beta-barrels. Due to these short length scales, coherent energy transfer becomes possible between the chromophores, and it has been shown that GFPs can sustain anomalously long-lived coherences between energetic states \-- with timescales of the order of picoseconds, many times longer than corresponding timescales in other bio-molecules~\cite{kim_venusa206_2019,CohDynGFP,FretGFP}, and that GFPs can undergo coherent energy transfer of excitations between adjacent chromophores at room temperature. Furthermore, the coherences have amplitudes measurable, even at room temperature, a striking result that cannot be accounted for using weak coupling theories, wherein Markov approximations are deployed. The timescales observed are commensurate with the dielectric relaxation in solvent water and it has been argued that the environment surrounding the chromophore - its beta-barrel and the solvent media - can allow for an extension of these timescales. However, at this time, a predictive framework capable of fully capturing all coherent energy transfer mechanisms and explaining these anomalously long timescales has not been available and the exploration and understanding of the precise physical mechanisms underpinning these phenomena is still an open area of research.

Early efforts on exploring decoherence in open quantum systems include the Caldeira-Leggett model for quantum Brownian motion~\cite{CALDEIRALeggett}, an elegant path integral approach, which unveiled mechanisms through which high-temperature Markovian environments can induce decoherence in quantum systems. Further studies have explored at length the archetypal quantum two-level system in dissipative environments~\cite{Leggett:Review}, but most of the approaches deployed are limited by their dependence on the Markovian approximation~\cite{Markov}, an approximation under which the evolution of  the system is entirely determined by its  present state. Markovian and weakly coupled system  dynamics can be described by generalized Lindblad master equations that usually take a simple form~\cite{IntroLindblad}; however, such an approximation is only valid if the environment in which the quantum system is placed relaxes on a timescale considerably shorter than the system dynamics. Hence, it does not capture the bidirectional flow of information between the system and environment: the system can only dissipate into the environment but does not have the ability to recapture previous excitations or correlations shared with the environment. Moreover, by accounting for non-Markovian effects~\cite{ReviewPaper}, the system's dynamics can be modulated drastically, increasing yield in chemical processes~\cite{BoostMolSwitch}, fractional decay in photonic crystals~\cite{SingleAtomSwitch}, as well as extending coherence lifetimes in multi-qubit systems~\cite{White2020} and theoretical models of energy transfer in photosynthetic systems~\cite{CoherencePS}.

While there are numerous techniques to model the non-Markovian dynamics induced by a structured environment, most often, these approaches require the deployment of a myriad of approximations in order to yield closed-form solutions or have tenable computation times.  In particular, for systems that are coupled only weakly to their environment, one can deploy perturbative approaches such as the second-order perturbation time-convolutionless master equation ~\cite{TheoryOQSBook}. Another approximation commonly utilised in the field of quantum optics is the rotating wave approximation, which neglects terms in the interaction Hamiltonian that have rapidly rotating phase contributions~\cite{StrongCouplingRWA}. Such an approximation is very helpful for low-temperature systems where the environmental frequencies are approximately equal to the transition frequency of a two-level system that is the system of interest and allows for the restriction of the Hilbert space into a subspace of a fixed excitation number~\cite{burgessNatom}.  

Notwithstanding, there exist a large number of systems wherein the approximations listed above are not valid. For example, for systems at moderate temperatures as well as intermediate coupling strengths, characterized by timescales overlapping between system and environment, not only the perturbation approximation, but also the RWA are impractical and can lead to anomalous results~\cite{ReviewPaper}. A better framework is therefore required to fully understand how decoherence occurs in these physical systems. In this work we employ the hierarchical equations of motion (HEOM)~\cite{HEOM,HEOMElecTransfer}, a non-perturbative, non-Markovian approach able to describe quantum systems coupled to structured environments on a wide range to temperature ranges by collecting dominant degrees of freedom in the environment and generating a hierarchy of equations from repeated time derivatives of the path integral.  To explore the physical mechanisms at the origin of the anomalous timescales in the dynamics of the GFPs, we extend the approaches of Gilmore \& McKenzie~\cite{McKenzieGilmore,Gilmore_2005,GILMORE2006266} and focus on the effects of dielectric relaxation on the coherence dynamics and energy transfer employing a spin-boson model for the GFP chromophore system. To capture the influence dielectric environments have, we employ the fluctuation and dissipation theorem~\cite{Kubo_1966} to derive a spectral density for the chromophore system and, by extension, its bath correlation function. We investigate  the decoherence-inducing phenomena in a HEOM framework for both an independent chromophore and a coherently coupled two-chromophore system and highlight the major shortcomings of  Markovian approach employing the Bloch-Redfield equations. Finally, we demonstrate  the non-Markovian negative entropy production by investigating the von Neumann entropy and its evolution under the time-dependent Gorini–Kossakowski–Sudarshan–Lindblad (GKSL) equation~\cite{Linblad}.

\subsection*{Spin boson model of chromophore systems}

The starting point in our exploration of the decoherence of chromophores in biological complexes by dielectric relaxation is that of a model system in which the chromophore is assumed to be a quantum point dipole in a quantized electromagnetic field. Two types of dipoles are present within such a system. Firstly, there are permanent dipoles generated by the anisotropy of the highest occupied molecular orbital (HOMO) and the lowest unoccupied molecular orbital (LUMO) of the chromophore~\cite{ColorMechanism}. Secondly, there are the purely quantum transition dipoles responsible for dissipation and excitation transfer via coupling to the electromagnetic field. However, due to the timescales of dephasing being on the order of ps, and radiative decay on the order of ns we can effectively neglect the transition dipoles when considering dynamical dephasing. As in Ref.~\cite{McKenzieGilmore}, we employ  a spin-boson model for the chromophore system by neglecting less relevant and slower-acting terms in the Hamiltonian. The system Hamiltonian takes the form
\begin{equation}
    H_1 =\frac{\omega_0}{2}\sigma_z + \sum_\lambda \omega_\lambda a^\dag_\lambda a_\lambda+ \Delta d\sigma_z \sum_\lambda g_\lambda(a_\lambda +a^\dag_\lambda),
    \label{eqn:reducedHam}
\end{equation}
where $\omega_0$ is the transition energy of the chromophore, $\omega_\lambda$ is the energy associated with the $\lambda$ mode of the electromagnetic field, $g_\lambda$ is the coupling strength of the difference dipole $\Delta \Vec{d}$, the difference between the permanent dipole in the HOMO and LUMO states $\Delta \Vec{d}= ({\Vec{d}_{H}-\Vec{d}_{L}})/({2})$, and the electromagnetic field. $\Delta d$ is the magnitude of the difference dipole. Here $a^{(\dag)}_\lambda$ denote the annihilation (creation) operator for the $\lambda$ mode of the electromagnetic field, and we work in a system of units such that $\hbar= 1$.  

For two spatially localized chromophores, their transition dipoles will interact via dipole-dipole interactions, enabling the F{\"o}rster resonant energy transfer and allowing for the direct, non-radiative exchange of excitations between the chromophores. Following the conventional approach, we focus on the single-excitation sub-space of the coupled chromophore systems~\cite{GILMORE2006266}, and consider the following single spin-boson Hamiltonian for the two chromophore system 
\begin{equation}
    H_2 = \frac{\Delta\omega}{2} \,\sigma_z + \Lambda\,\sigma_x + \,\sigma_zV + \sum_\lambda \omega_{1\lambda} a^\dag_{1\lambda} a_{1\lambda}+ \sum_\lambda \omega_{2\lambda} a^\dag_{2\lambda} a_{2\lambda},
    \label{eqn:SBHomoDimer}
\end{equation}
with $\Delta\omega = \omega_1-\omega_2$  the difference between the two chromophore transition frequencies and where the operator
\begin{align}
    V &= \Delta d_1 \,\sum_\lambda g_{1\lambda}(a_{1\lambda}+a^\dag_{1\lambda}) - \Delta d_2 \, \sum_\kappa g_{2\kappa}(a_{2\kappa}+a^\dag_{2\kappa})
\end{align}
captures the coupling of each chromophore to its local electromagnetic field environment. 
Here 
\begin{equation}
    \Lambda = \frac{\kappa d_1 d_2}{n^2R^3\varepsilon_0},
\end{equation}
with $\kappa$ is the dipole orientation factor, $d_{1,2}$  the transition dipoles associated with each chromophore, $n$ the refractive index of the medium between the two chromophores, $R$ the relative distance between them and the index $i$ labelling each chromophore and its local environment.  

The problem is now equivalent to a single spin-boson model with a tunnelling term ($\propto \Lambda$) between the two singly excited states of the two chromophores  $\ket{1}=\ket{e}_1\otimes\ket{g}_2$ and $\ket{0}=\ket{g}_1\otimes\ket{e}_2$. If we assume the two environments are uncorrelated due to a suitable spatial separation, the spectral density associated with the system is simply the sum over the individual chromophore spectral densities: $ J(\omega) = J_1(\omega)+J_2(\omega).$

\begin{figure}[ht]
    \centering
    \begin{subfigure}{0.42\textwidth}
    \centering
    \includegraphics[width=\textwidth]{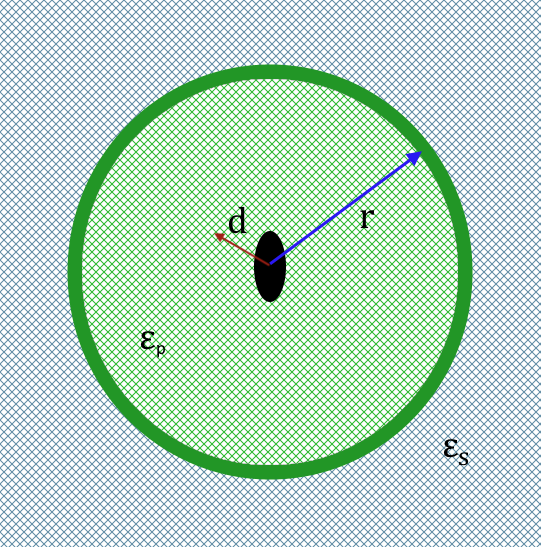}
    \caption{}
    \label{fig:Model1}
    \end{subfigure}
    \begin{subfigure}{0.42\textwidth}
    \centering
    \includegraphics[width=\textwidth]{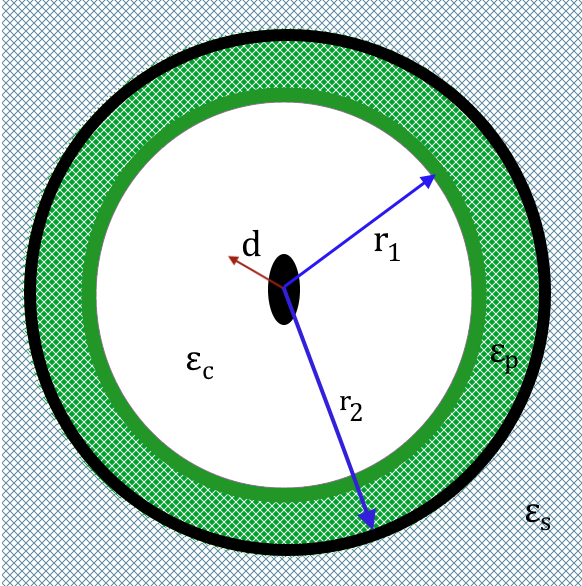}
    \caption{}
    \label{fig:Model2}
    \end{subfigure}
    \caption{System architectures for models of chromophores given by dipole $d$ embedded within (a) protein complex with dielectric given by $\varepsilon_\textrm{p}$ with radius $r$ inside of solvent medium with dielectric $\varepsilon_\textrm{s}$. (b) a cavity of radius $r_1$ with dielectric $\varepsilon_\textrm{c}$ in a protein complex of radius of radius $r_2$ and dielectric $\varepsilon_\textrm{p}$ inside of solvent medium with dielectric $\varepsilon_\textrm{s}$. }
    \label{fig:Models}
\end{figure}

 This formalism provides the generic spin-boson model for exploring dephasing phenomena due to dielectric relaxation. Such models have been studied extensively, but previous studies were within a limited regime characterized by weak coupling or Markovian approximations imposed upon the dynamics. To study the system's non-Markovian dynamics from the above-derived Hamiltonians,  we need to identify a spectral density model that characterises the structure associated with the local environment. Here, we use the approach from Refs.~\cite{McKenzieGilmore,Gilmore_2005} to generate spectral densities for the system architectures shown in Fig.~\ref{fig:Model1} and \ref{fig:Model2}. In Fig.~\ref{fig:Model1},  we consider a simplistic geometry representing a chromophore embedded inside a spherical protein network within a solvent medium, whereas in \ref{fig:Model2} we consider the same architecture with the addition of a cavity within which the chromophore is embedded. Subsequently, we generate spectral densities associated with these models by characterising the dielectric natures of each and applying the Fluctuation and Dissipation Theorem. 

In principle, all the dielectrics in each region of the models are frequency dependent. However, due to the small extent of the protein, variations in the solvent environment will be the most critical to the dynamics. For the standard solvent - water - we can characterize the low-frequency dielectric spectral region using a Debye model, with a frequency-dependent dielectric permittivity of  the form
\begin{equation}
    \varepsilon(\omega) = \varepsilon_\infty + \frac{\varepsilon_\textrm{s}-\varepsilon_\infty}{1+i\omega \tau_D}.
\end{equation}
$\varepsilon_\infty$ and $\varepsilon_\textrm{s}$ are the high and low-frequency limit dielectrics for the solvent, respectively. For both system architectures considered in Fig.~\ref{fig:Models},  this results in Ohmic spectral densities with a Lorentz-Drude cutoff of the form,
\begin{equation}
    J(\omega) = \frac{2\omega\lambda \gamma}{\pi(\omega^2 + \gamma^2)},
\end{equation}
where $\lambda$ is the coupling strength and $\gamma$ the cut-off frequency. 

\bigskip

From the Fluctuation and Dissipation theorem, we can relate the zero-temperature response to the finite temperature response through the correlation function

\begin{equation}
    C(t) =\int^\infty_{0} d\omega \, J(\omega)\left[\text{coth}(\omega/2T)\,\text{cos}(\omega\tau)-i\, \text{sin}(\omega \tau) \right].
\end{equation}

In order to determine the correlation function for this spectral density, we need to perform the Matsubara decomposition of the $\text{coth}$ term such that 
\begin{equation}
     \text{coth}\left(\frac{\omega}{2T}\right) = 2T\left(\frac{1}{\omega}+ \sum_{k=1}^\infty \frac{2\omega}{\omega^2 +(2\pi k T)^2} \right).
\end{equation}
This allows for a term-by-term integration of the integral appearing in the correlation function, which yields a natural decomposition of the correlation function
\begin{equation}
    C(t) = \sum^\infty_{k=0} c_k e^{-\nu_kt},
\end{equation}
with the frequencies given by 
\begin{equation}
    \nu_k = \begin{cases}
               \gamma               & k = 0\\
               {2 \pi k T}  & k \geq 1\\
           \end{cases},
\end{equation}
and the amplitudes by 
\begin{equation}
     c_k = \begin{cases}
               \lambda \gamma (\cot( \gamma / 2T) - i)             & k = 0\\
               4 \lambda \gamma \nu_k T/ (\nu_k^2 - \gamma^2)  & k \geq 1\\
           \end{cases}.
\end{equation}

However, as argued above, this single Debye timescale only describes the dielectric’s permittivity at low frequencies. To spectral responses in the optical regime - relevant to the bio-molecular complexes - we need to introduce a secondary timescale,  associate with  movement of defects in the hydrogen bond network~\cite{DDebye}. In this regime, the dielectric for water can be well described by a double Debye model given by
\begin{equation}
    \varepsilon(\omega) = \varepsilon_\infty +\frac{\varepsilon_\textrm{s} -\varepsilon_1}{1+i\omega \tau_D} + \frac{\varepsilon_1 - \varepsilon_\infty}{1+i\omega \tau_1}.
\end{equation}

Such a dielectric generates spectral densities of the form
\begin{equation}
    J(\omega) = \lambda\frac{a\omega + b\omega^3}{c+d\omega^2+f\omega^4},
    \label{eqn:SpecDensDD}
\end{equation}
with the explicit parametrization of these variables given in Appendix.~A. Due to the even nature of the integrand, we extend the integral between $[-\infty,\infty]$ such that 
\begin{equation}
    C(t) =\frac{1}{2}\int^\infty_{-\infty} d\omega \, J(\omega)\left[\text{coth}(\omega/2T)\,\text{cos}(\omega\tau)-i\, \text{sin}(\omega \tau) \right].
\end{equation}
Considering now the contour integral in the upper half plane, we need only calculate the poles associated with the spectral density factor and the $\coth$ term. For the temperature-dependant $\coth$ term, the poles are simply at the Matsubara frequencies, $\omega_k = \pm 2\pi i k T$ . Similarly, the poles in the spectral density occur at
\begin{equation}
    \omega = \pm\sqrt{\frac{-d\pm\sqrt{d^2-4cf}}{2f}}.
\end{equation}

Altogether, the bath correlation function can be written as
\begin{align} 
C(t) =& \frac{\lambda\pi}{2\zeta}\sum_{j=+,-} -j(a-b\alpha_j^2)e^{-\alpha_jt}\left[ \text{cot}\left(\frac{\alpha_j}{2T}\right) -i\right]\nonumber
    \\
    &
    +\sum_{n=1}^\infty 2\pi\lambda T\frac{e^{-\omega_n t}\omega_n(-a+b\omega_n^2)}{c-d\omega_n^2+f\omega_n^4},
\end{align}
where $\zeta = \sqrt{d^2-4cf}$ and $\alpha_\pm^2 =\dfrac{d\pm\zeta}{2f}$ are the poles associated with the spectral density.
This can then be recast as the standard series representation of the correlation function 
\begin{equation}
    C(t) = \sum_{k=-1}^\infty c_ke^{-\nu_kt},
\end{equation}
with the frequencies given by 
\begin{equation}
    \nu_k = \begin{cases}
               \alpha_-               & k = -1\\
               \alpha_+               & k = 0\\
               {2 \pi k}T & k \geq 1\\
           \end{cases},
\end{equation}
and the amplitudes by 
\begin{equation}
     c_k = \begin{cases}
              \frac{\lambda\pi}{2\kappa}(a-b\alpha_-^2)\left[ \text{cot}\left(\frac{\alpha_-}{2T}\right) -i\right]             & k = -1\\
              \frac{\lambda\pi}{2\kappa}(b\alpha_+^2-a)\left[ \text{cot}\left(\frac{\alpha_+}{2T}\right) -i\right]            & k = 0\\
               2\pi\lambda T\frac{\nu_k(-a+b\nu_k^2)}{c-d\nu_k^2+f\nu_k^4}   & k \geq 1\\
           \end{cases}.
\end{equation}
 
 This form of the correlation function makes it possible to perform a direct comparison between the single and the double Debye models and considerably simplifies the identification of the relevant time scales in the problem. We also note that the thermal fluctuations in the environment produce additional timescales in the system's dynamics, precisely at one over each of the Matsubara frequencies. Furthermore, the spectral density of the double Debye model has a super-Ohmic character for higher energies signalling that compared to the single Debye model, the system will interact stronger with electromagnetic modes located at higher frequencies in the electromagnetic spectrum.
  
\section*{Non-Markovian Dynamics in Green Fluorescent Proteins}\label{sec2}

 Due to the spectral structuring of the electromagnetic reservoir,  we expect the environment to provide non-negligible information feedback to the system which can directly affect the temporal scales over which non-vanishing coherences are present in the system. To fully capture these mechanisms we deploy the HEOM~\cite{HEOM,HEOMElecTransfer} formalism to explore the dynamics of the single and dimeric chromophore systems. The HEOM formalism (see Methods section) is  a non-perturbative, non-Markovian approach able to describe quantum systems coupled to structured environments on a wide range of temperatures. 
  
\begin{figure}
    \centering
    \includegraphics[width=0.5\textwidth]{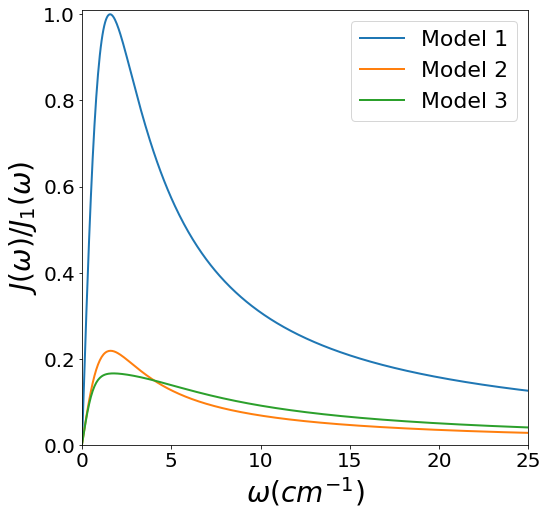}
    \caption{Spectral density associated with a GFP under different approximations of its structure normalised with respect to Model 1. Model 1: a simple protein solvent environment described by a single Debye timescale as in Fig.~\ref{fig:Model1}. Model 2: for a cavity protein solvent environment with a single Debye timescale as in Fig.~\ref{fig:Model2}. Model 3:  for a cavity protein solvent environment with a double Debye timescale as in Fig.~\ref{fig:Model2}.}
    \label{fig:SpectralD}
\end{figure}

 The model system we consider is that of the GFP,  a system that is of major interest as it has been experimentally suggested that its decoherence timescale is anomalously long compared to other biomolecules, on the order of ps. Furthermore, the cylindrical shape of the $\beta$-barrel surrounding the small chromophore allows for compatibility with the structural architectures studied in Models 1, 2 and 3. The chromophore in GFP has a transition energy at $490$nm, yielding its distinct green colouration.  Figure~\ref{fig:SpectralD} shows the spectral densities of the three models we consider in this study using system parameters commonly available when classifying green fluorescent proteins, given in Table.\ref{table}: model 1: a simple protein solvent environment described by a single Debye timescale as in Fig.~\ref{fig:Model1}; model 2: for a cavity protein solvent environment with a single Debye timescale as in Fig.~\ref{fig:Model2}; model 3:  for a cavity protein solvent environment with a double Debye timescale as in Fig.~\ref{fig:Model2}. These parameters generate effective timescales for the spectral densities ($\gamma^{-1}$, $\alpha_\pm^{-1}$) of $0.63$ps for model 1, $0.62$ps for model 2 and $0.99$ and $0.24$ps for model 3 that has two effective timescales, due to the double Debye model. Clearly, for model 1, due to the chromophore being in direct contact with the protein, the coupling is significantly stronger compared to models 2 and 3, wherein the chromophore is embedded within a cavity of a significantly lower dielectric. Here, models 2 and 3 have the same system geometry, but different numbers of Debye timescales used to model the dielectric nature of the solvent; model 3 with two Debye timescales has lower coupling strength for small frequencies, compared to model 2, but for higher frequencies, this relationship is inverted, due to the super-Ohmic nature of the spectral density.  

To fully understand to what extent the non-Markovian effect influences the system's dynamics, we also present a purely Markovian set of results and contrast them with the results produced by the HEOM method. For the Markovian approach, we adopt the Bloch-Redfield formalism, which is equivalent to a second-order perturbation master equation with an additional Markov approximation of the second kind and removal of energy Lamb shifts induced by the environment~\cite{REDFIELD19651,TheoryOQSBook}  (see Methods section).

\subsubsection*{Single Chromophore System}

We start by exploring the dynamical decoherence of the two-level chromophore system inside of a GFP environment. To unveil the coherence dynamics, we choose as an initial condition with  the chromophore in a superposition state between the ground and excited state, with an initial density operator given by
\begin{equation}
    \rho(0) = \frac{1}{2}\left(\ket{1}+\ket{0}\right)\left(\bra{1}+\bra{0}\right)\otimes \rho_E^\beta .
\end{equation}

\begin{figure*}[ht]
    \centering
    \begin{subfigure}{0.4\textwidth}
    \centering
    \includegraphics[width=\textwidth]{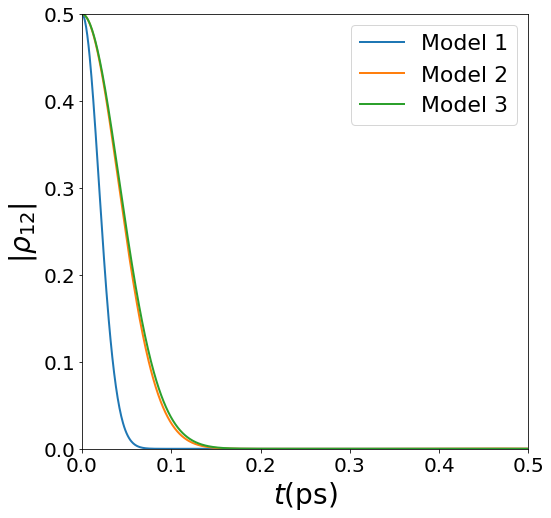}
    \caption{}
    \label{fig:SingleHEOM}
    \end{subfigure}
    \centering
    \begin{subfigure}{0.4\textwidth}
    \centering
    \includegraphics[width=\textwidth]{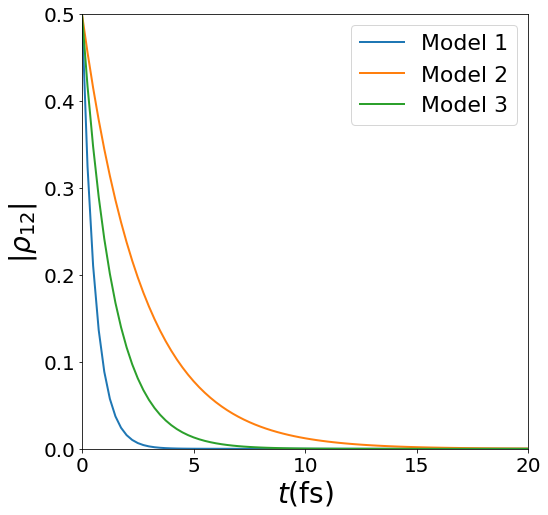}
    \caption{}
    \label{fig:SingleBR}
    \end{subfigure}
    \caption{Dynamical decoherence of the two-level chromophore within GFPs using (a) the non-Markovian Hierarchical equations of motion (b) the Markovian Bloch-Redfield equations. The various models used are, Model 1: a simple protein solvent environment described by a single Debye timescale as in Fig.~\ref{fig:Model1} Model 2: for a cavity protein solvent environment with a single Debye timescale as in Fig.~\ref{fig:Model2} Model 3:  for a cavity protein solvent environment with a double Debye timescale as in Fig.~\ref{fig:Model2}. The parameters were taken from the literature and are given in Table~\ref{table}.}
    \label{fig:SingleGFP}
\end{figure*}
Figure~\ref{fig:SingleGFP} contrasts the excitation dynamics of single chromophore structure under Markovian and non-Markovian approximations. The exact dynamics of the system are sensitive to the model used for the surrounding environment. In Model 1, the decay is considerably faster since the system is now in direct contact with the protein network leading to a stronger coupling as evidenced by the spectral densities shown in Fig.~\ref{fig:SpectralD}. However, both Model 2 and 3 appear to have very similar dynamics in the HEOM regime. This is because both system and interaction Hamiltonians are proportional to $\,\sigma_z$, which is not modulated in the interaction picture; hence, the low-frequency environmental degrees of freedom play the dominant role in the dynamics of the system. As shown in Fig.~\ref{fig:SpectralD}, these two models have nearly identical spectral densities at low frequencies, accounting for their similar behaviour. We also note the sub-exponential decay of the coherence in the system in Fig.~\ref{fig:SingleHEOM}; at early times, the decay rate is significantly less before saturating at later times. This time-dependent decay rate is a key characteristic of non-Markovian dynamics. Our results also predict that dielectric relaxation effects occur up to almost picosecond timescales. This brings the theoretical modelling and experimental results data for the single GFP closer together~\cite{CohDynGFP,FretGFP,KIMVenus} (also ps timescales)  and demonstrates the efficacy of the HEOM approach to understanding the dynamical features of decoherence in these bio-molecular systems. Moreover, our results suggest that the solvent media dominates the decoherence rather than the specific protein dynamics. One would expect the specific frequency of oscillations to be determined by the nature of the vibrational environment, but the solvent environment dominates the decoherence timescale.
 
The Bloch-Redfield results shown in Fig.~\ref{fig:SingleBR}, clearly demonstrate that the Markovian approximation predicts a much shorter decoherence times scale, hence the longer coherence times obtained in the HEOM formalism are associated with memory effects induced by information backflow from the environment to the system: under the non-Markovian system description, the decoherence dynamics is dominated by ps time scales, whereas the Bloch-Redfield approach gives rise to time scales of the fs order. The failure of the Bloch-Redfield equations is apparent as we note that the most natural system timescale, associated with the transition frequency $\omega_0$, is significantly faster than the timescales associated with the environment as $\omega_0/\gamma \approx 2000$. This contradicts the necessary condition for the validity of the Markov approximation that underpins the Bloch-Redfield equations. Physically, we are demanding that the environment relaxes instantaneously and provides only a small detriment to the system coherences. However, by allowing the environment to relax over an effective timescale commensurate with the system's dynamics, we have an extension of the chromophore's coherence lifetime. This can be traced back to the environment correlation functions in the Bloch-Redfield theory. The decay rates saturate immediately are immediate, equivalent to rapid saturation of the correlation functions resulting in delta functions and, thus allowing no temporal correlations. In contrast, in the entire system HEOM calculation, the correlation function decays away at timescales of ps, allowing for suppression of the dephasing.

Further insight can be gained by analysing the analytical solution of the Bloch-Redfield equations for $\rho_{12}$:
\begin{equation}
    \rho_{12} = \rho_{12}(0)e^{-(2\Gamma+i\omega_0)t},
\end{equation}
with the rate constant
\begin{equation}
    \Gamma = 2\lambda T/\gamma,
\end{equation}
for models 1 and 2. Clearly, then dynamics of the $\mathbf{l}_1$ norm coherence measure~\cite{PlenioCoh}
\begin{equation}
\mathbf{C} = 2\abs{\rho_{12}} = 2\abs{\rho_{12}(0)}\,e^{-2\Gamma t}
\end{equation}
is simply an exponential decay. Since the decay rate $\Gamma$ is directly proportional to the coupling strength $\lambda$, we can easily explain the different dynamics displayed in Fig.~\ref{fig:SingleBR}: Model 1 decays the fastest due to its higher coupling strength. However, this coupling strength dependence is not as stark in the full non-Markovian dynamics displayed in Fig.~\ref{fig:SingleGFP} as Models 2 and 3 express very similar dynamics.

\subsubsection*{Homo-Dimer Energy Transfer}

We now deal with a more exciting and challenging topic and consider the problem of coherent energy transfer between two-like GFPs, which can non-radiatively transfer excitations through dipolar interactions between the transition dipoles of both chromophores. We employ the Hamiltonian in Eq. \ref{eqn:SBHomoDimer}, which for a homo-dimer system has site transition energy $\omega = 0$ . Here, we consider an initial state in which the first chromophore is in its ground state and the second is in its excited state. In the reduced single spin subspace, this can be expressed as $ \rho = \ket{1}\bra{1}\otimes\rho_B^\beta$.
\begin{figure*}[ht]
    \centering
    \begin{subfigure}{0.45\textwidth}
    \centering
    \includegraphics[width=\textwidth]{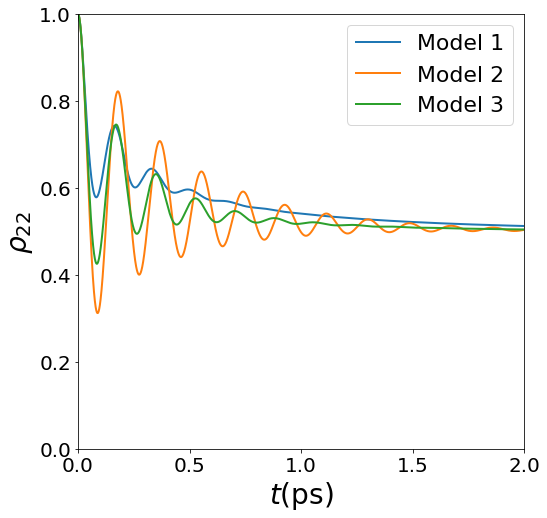}
    \caption{}
    \label{fig:DimerHEOM}
    \end{subfigure}
    \begin{subfigure}{0.45\textwidth}
    \centering
    \includegraphics[width=\textwidth]{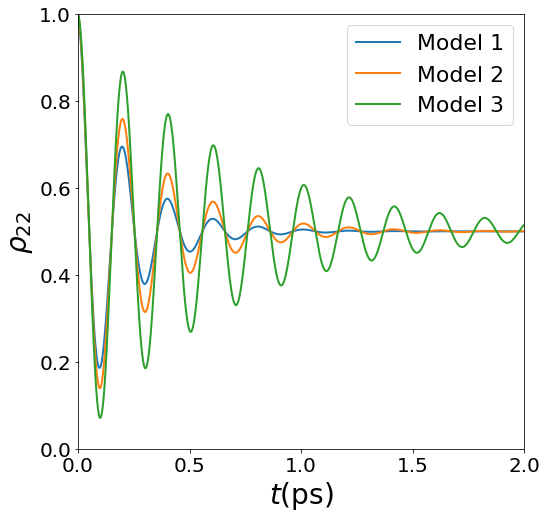}
    \caption{}
    \label{fig:DimericBR}
    \end{subfigure}
    \caption{Excitation dynamics of two-coupled chromophores within GFPs using (a) the non-Markovian HEOM formalism (b) the Markovian Bloch-Redfield equations. The models used are shown in Figs.~\ref{fig:Model1}-\ref{fig:Model2}. The parameters used are given in Table~\ref{table}.}
    \label{fig:HomoDimerGFP}
\end{figure*}

The homo-dimer GFP system excitation dynamics (see  Fig.~\ref{fig:HomoDimerGFP}) is considerably richer than the one obtained in the monomeric case. Firstly, we note that in each of the models used, coherent energy transfer occurs due to the oscillations in the probability of the excitation being localized in the second chromophore $\rho_{22}$. These contrast with the non-oscillatory exponential decay associated with incoherent energy transfer. These oscillations occur much faster than the decoherence rate, allowing the excitation to be coherently transferred between the chromophores. It is worth recalling that the only difference between the models used is the associated spectral density, while the dipole-dipole interactions are identical. Hence, the modulations in the energy transfer rates - the periodicity of the oscillations - are purely induced by the interaction with the environment.

In the Bloch-Redfield regime, we note that the dynamics are qualitatively similar to the HEOM results. However, in the non-Markovian HEOM case, the oscillations are not about $\rho_{22}=\frac{1}{2}$ as in the Bloch-Redfield results; this is due to the Bloch-Redfield equation neglecting Lamb shifts in the systems. These Lamb shifts introduce an asymmetry in the transition frequencies between the two chromophores, suppressing the oscillations of the energy transfer, effectively making it more difficult to transfer energy. We also note that whilst the Model 3 system has a very long coherence lifetime in the Bloch-Redfield regime, in the HEOM results, we see that this timescale is shorter than that associated with Model 2. This is due to the Bloch-Redfield solutions simply sampling the spectral density at the tunnelling rate $\Delta$, as we will show shortly, whereas, in the HEOM regime, nearby frequencies to the tunnelling rate will also play a key role.

\begin{figure}
    \centering
    \includegraphics[height=8cm]{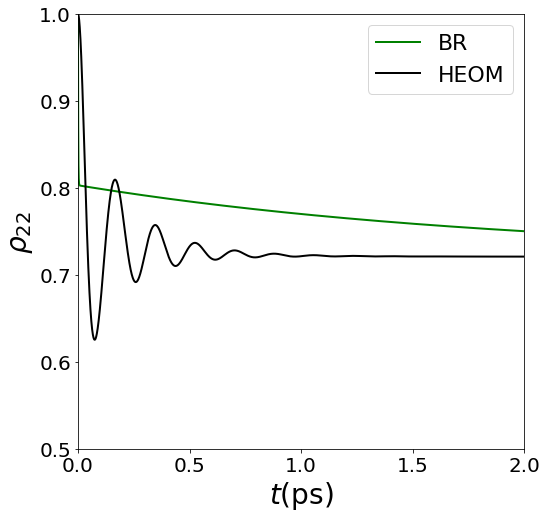}
    \caption{The dynamics of the population dynamics of a chromophore in a Homo-Dimer of GFP, comparing the predictions of the Bloch-Redfield theory to the HEOM approach  for a small asymmetry in the transition frequencies of the two chromophores: $\Delta\omega = \omega_1-\omega_2 = 0.01\omega_1$ }
    \label{fig:asym}
\end{figure}

We again consider the exact solutions for the homo-dimer Bloch-Redfield equations, with the population of the first chromophore given by
\begin{equation}
    \rho_{11} =\frac{e^{-\gamma_0t}}{2\Gamma}\left(\gamma_0 \sinh{\Gamma t}+ \Gamma \cosh{\Gamma t} \right) + \frac{1}{2},
\end{equation}
where the two effective rates are 
\begin{equation}
    \Gamma = \sqrt{\gamma_0^2-4\Lambda^2},
\end{equation}
and 
\begin{equation}
    \gamma_0 = 8\pi J(2\Lambda)\coth{(\beta \Lambda)}.
\end{equation}
As shown in Fig.~\ref{fig:DimericBR}, the $\Gamma$  rate is imaginary hence inducing coherent oscillations. However, the $\gamma_0$ rate is linearly dependent on the coupling strength $\lambda$ of the spectral density through $J(2\Lambda)$. This is not the case in the non-Markovian dynamics displayed in Fig.~\ref{fig:HomoDimerGFP},  where in Model 1 the chromophore couples significantly stronger to the local electromagnetic field than in the other models. Our results show that the inter-chromophore coupling rate is similar to the coupling strength with $\Lambda/\lambda \approx 3$ for Models 2 and 3. In this regime, the system's dynamics occur on timescales commensurate with the reorganization of the environment, which is not captured in the Markovian approximation, which asserts that the environment relaxes infinitely quickly. We can easily explore the difference between the effectiveness of the Bloch-Redfield solutions in the dimer case and the single GFP case. In the dimer case, the chromophores have the same energy, so the inter-site exchange requires no additional energy cost. The only relevant energy scale is associated with the inter-chromophore coupling. This is mediated by dipole-dipole interactions, a weak force, and is considerably smaller than the energy gap in the single chromophore system with ${\omega_0}/{\Lambda}\approx300$. As such, the timescales in the single chromophore system are much faster, pushing the dynamics into a more non-Markovian regime, necessitating the HEOM approach. Considering a more physical situation in which (due to local asymmetries) there is a non-zero difference between the transition energies of the two chromophore $\Delta\omega = \omega_1-\omega_2\neq 0$, we find that the Bloch-Redfield equations are much less effective. The results presented in Fig.~\ref{fig:asym} show that even for a minimal asymmetry ($\Delta\omega = \omega_1-\omega_2= 0.01\omega_1$) that the Bloch-Redfield theory is inadequate at capturing the system dynamics. This also suggests it would become impractical in the case of hetero-dimer systems, where the energy differences are considerably larger.
Furthermore, we note that, in contrast, the HEOM results are robust to this type of perturbation.

\begin{figure}
    \centering
    \includegraphics[height=8cm]{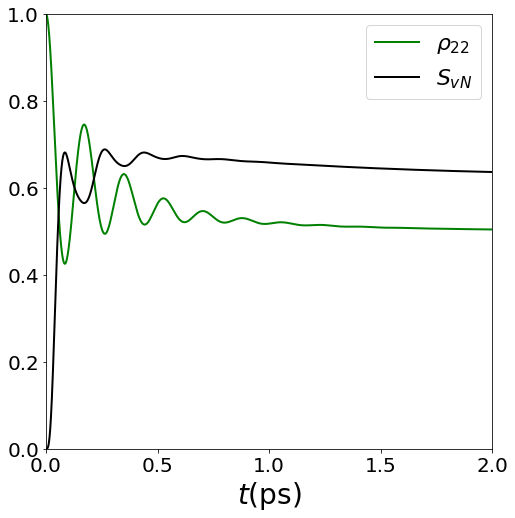}
    \caption{The population and entropy dynamics of a homo-dimer GFP in Model 3: for a cavity protein solvent environment with a double Debye timescale.}
    \label{fig:Entropy}
\end{figure}

We note that for both the single chromophore and the homo-dimer systems, the Bloch-Redfield equations are insufficient at capturing the full dynamics as relevant rates in the system are directly proportional to the coupling strength $\lambda$. Another unexpected aspect of the dynamics of these models is the transient lowering of the von Neumann entropy, as shown in Fig.~\ref{fig:Entropy}, often associated with memory effects present in the system. Additionally, there is a direct correlation between the oscillations due to coherent energy transfer and the reduction of the von Neumann entropy. We argue that precisely this interplay allows for an extension of the decoherence of the system compared to the single system, which we explore next.

\subsubsection*{Negative Entropy Production}
To fully explore the implications of the complex dynamics described above, we focus now on the evolution of the system's entropy. To this end, we consider the von Neumann entropy of a density matrix $\rho$ given by 
\begin{equation}
    S_{vN} = -\Tr{\rho \ln{\rho}}.
\end{equation}
The entropy production can be found by taking the time derivative 
\begin{equation}
    S_{vN} = -\Tr{\Dot{\rho} \ln{\rho}}.
\end{equation}
Using the HEOM-Gorini–Kossakowski–Sudarshan–Lindblad (GSKL) formalism mapping presented in the Methods Section, we can rewrite the entropy production as 
\begin{align*}
     \Dot{S}_{vN} &=- \Tr{\ln{\rho}\sum_{\alpha,\beta}\Gamma_{\alpha\beta }\left( \,\sigma_{\alpha} \rho \,\sigma_{\beta}^\dag -\frac{1}{2}\{\,\sigma_{\beta}^\dag \,\sigma_{\alpha},\rho\}\right)}\\
     &=- \sum_{\alpha,\beta}\Gamma_{\alpha\beta }\Tr{\ln{\rho}\left( \,\sigma_{\alpha} \rho \,\sigma_{\beta}^\dag -\rho\,\sigma_{\beta}^\dag \,\sigma_{\alpha}\right)}.
\end{align*}

Assuming that the cross terms in the previous equation can be neglected, the double sum reduces to a single sum. Then taking the operators $\{\,\sigma_\alpha\}$ to be the set of generalized Pauli operators (Hermitian and involutions, $\,\sigma_\alpha^2 =\mathbf{1} $), we obtain
\begin{align}
    \Dot{S}_{vN}& =-\sum_{\alpha}\Gamma_{\alpha}\Tr{\ln{\rho}\left( \,\sigma_{\alpha} \rho \,\sigma_{\alpha} -\rho\right)}\\
    &=\sum_{\alpha}\Gamma_{\alpha}\left(\Tr{\rho\ln{\rho}} -\Tr{\Tilde{\rho}_\alpha \ln{\rho}}\right)\\
    &= \sum_{\alpha}\Gamma_{\alpha}S(\Tilde{\rho}_\alpha{\rvert\rvert}\rho)
\end{align}
Here $\Tilde{\rho}_\alpha = \,\sigma_\alpha\rho\,\sigma_\alpha$ is the basis transformation of the system state to that associated with eigenstates of the Lindblad operator $\,\sigma_\alpha$ and $S(\Tilde{\rho}_\alpha{\rvert\rvert}\rho)$ is the relative entropy of the density matrix coordinate transformed to the basis of $\,\sigma_\alpha$ with the untransformed density matrix. 

Consequently, we expect the relative entropy to be non-negative valued and zero only when $\rho_\alpha = \rho$. Therefore, for negative entropy production, we require that both rates $\Gamma_\alpha$, take negative values. These negative valued Lindblad rates are associated with a flow of energy or information from the environment into the system~\cite{TheoryOQSBook}, which is exemplary of non-Markovian dynamics~\cite{NMEntropy}. These negative entropy production events can also manifest as a generation of coherence in the quantum system~\cite{Wang2017}. 

\section*{Discussion}

To conclude, we have developed a framework to describe bio-molecular decoherence inside GFPs using a reduced spin-boson model. Our approach deals with the decoherence mechanisms induced by dielectric relaxation and predicts timescales of similar magnitude to those measured experimentally~\cite{CohDynGFP,FretGFP,KIMVenus} and employs a reservoir spectral density derived directly from the fluctuation-dissipation theorem applied to the local electromagnetic field surrounding the chromophore. It fully captures the influence of dielectric properties of the environment on the chromophore dynamic and demonstrates the significant influence associated with  different environmental configurations strongly on the  dynamical decoherence of the GP complexes. Furthermore, we have shown the non-Markovian bi-directional flow of information between the system and dielectric environment plays an essential role in the system dynamics. We have also shown that certain configurations (single chromophore or the slightly asymmetric homo-dimer system) can lead to long decoherence timescales. Our exploration of the von Neumann entropy evolution strongly supports the argument that the transient fluctuations in the entropy of the reduced system and the associated negative entropy production are also linked to non-Markovian effects. The formalism developed here can also be deployed to investigate other complex bio-molecular systems undergoing coherent energy transfer, including hetero-dimers, trimers, tetramers and light-harvesting complexes.

 \printbibliography[heading=subbibliography]
 
 \end{refsection}

 \begin{refsection}
\section*{Methods}\label{sec11}

\subsection*{Hierarchical Equations of Motion}
The hierarchical equations of motion (HEOM) method utilizes a discretization of the environment to generate a numerically efficient approach to capturing the effects of the collective degrees of freedom of the environment. This is achieved by constructing coupled differential equations that form a hierarchy from repeated time-derivatives of the influence functional, derived initially by Vernon and Feynman~\cite{FEYNMAN1963118}. Such an approach is non-perturbative and allows for the exploration of strongly coupled systems with a non-zero temperature environment~\cite{HEOM}. A fundamental assumption in this approach is that the system couples linear to its environment and that the bath correlation functions can be conveniently decomposed in a Fourier series. For a generic environment spectral density, this decomposition will be non-trivial; as such, we may adopt a numerical approach by fitting with spectral densities that do have convenient decompositions.\par
Another limitation of the HEOM method is the absence of a general approach in deciding how to truncate the hierarchy - which is infinite for a bosonic environment - and the Fourier decomposition of the correlation functions. The truncation is dependent on the initial temperature of the environment as well as the complexity of the spectral density associated with it.\par

In this work, we are specifically interested in bosonic reservoirs and their influence on the two-level systems. This is because we are modelling the electromagnetic field - a bosonic reservoir - and its interactions with the quantum dipole - a two-level system - of the chromophore. The Hamiltonian for a system interacting with a bosonic environment utilizing the second quantization is simply
\begin{equation}
    H = H_S  + \sum_\lambda \omega_\lambda a^\dag_\lambda a_\lambda + Q \sum_\lambda g_\lambda(a_\lambda + a^\dag_\lambda). 
\end{equation}
where $H_S$ is the free system Hamiltonian, and $Q$ is the coupling operator in the system's degrees of freedom. 
For convenience,  we introduce the bath operator
\begin{equation}
    E = \sum_\lambda g_\lambda(a_\lambda + a_\lambda^\dag).
\end{equation}
By utilizing the Feynman-Vernon influence functional approach to path integrals, we can define the time evolution of the reduced density matrix as 
 \begin{align}
    \Tilde{\rho}_S(t) =& \mathcal{T}\text{exp}\Biggl\{-\int^t_0 dt_2\int^{t_2}_0dt_1 Q^\times (t_2)\biggl[C_R(t_2-t_1)Q^\times(t_1) \nonumber\\& +\,i\, C_I(t_2-t_1)Q^o(t_1)\biggr]\Biggr\}\rho_S(0)
    \label{eqn:FVIFrho}
\end{align}
 
The tilde on the density operator signals that we are working in the interaction picture at time $t$, such that operators transform under the rule $A\rightarrow e^{i(H_S+H_E)t}Ae^{-i(H_S+H_E)t}$, where  $H_B = \sum_\lambda \omega_\lambda a^\dag_\lambda a_\lambda$ is the free environment Hamiltonian. We also assume that the initial state of the entire system is unentangled: $\rho(0)=\rho_S(0)\otimes \rho_B$ where $\rho_B$ is the thermalized Gibbs state given by
\begin{equation}
    \rho_B = \frac{e^{-\beta H_B}}{Z},
\end{equation}
where the inverse temperature of the reservoir $\beta = 1/T$ and $Z = \text{tr}\{e^{-\beta H_B}\}$. Here, we have also introduced the notation for the commutator and the anticommutator
\begin{equation}
    Q^\times = [Q,\cdot] \text{, and } Q^0 = \{Q,\cdot\}.
\end{equation}
Due to the harmonic nature of the environment, we can assume the environment acts as a Gaussian noise source, as described in the fluctuation-dissipation theorem. Under this assumption, the induced dynamics depend only on the second-order correlations of the environment defined by
\begin{align}
    C(\tau) = \langle E(t+\tau)E(t)\rangle = \int^\infty_0 d\omega J(\omega)\left[\text{coth}(\beta\omega/2)\text{cos}(\omega\tau)-i\text{sin}(\omega \tau) \right].
\end{align}
To facilitate the introduction of a hierarchical set of equations of motion, we decompose these correlation functions into real and imaginary parts
\begin{equation}
    C(t) = C_R(t)+iC_I(t),
\end{equation}
and corresponding Fourier components  
\begin{equation}
    C_R(t) = \sum^{N_R}_{k=1}c_k^R e^{-\gamma_k^Rt},
\end{equation}
\begin{equation}
    C_I(t) = \sum^{N_I}_{k=1}c_k^I e^{-\gamma_k^It},
\end{equation}
where the coefficients $c_k^{R,I}$ and frequencies $\gamma^{R,I}_k$ can, in principle,  be complex-valued. Applying consecutive time derivatives on the time evolution of the reduced density matrix described in Eq.~\ref{eqn:FVIFrho},  we generate an infinite set of coupled first-order equations of the form
\begin{align}
    \Dot{\rho}^n(t) &= \Biggl( -iH^\times_S-\sum_{j=R,I}\sum_{k=1}^{N_j}n_{jk}\gamma^j_k \Biggr)\rho^n(t) \nonumber \\ &-i\sum^{N_R}_{k=1}c_k^R n_{Rk}Q^\times \rho^{n^-_{Rk}}(t) + \sum^{N_I}_{k=1}c_k^In_{Ik}Q^o\rho^{n^-_{Ik}}(t) \nonumber \\
    &-i\sum_{j=R,I}\sum_{k=1}^{N_j}Q^\times \rho^{n^+_{jk}}(t),
\end{align}
where we have introduced the multi-index $n = (n_{Ri},n_{Ii}), n_{R,Ii}\in \{0,...,N_c\}$ where $N_c$ is the cutoff parameter defining the depth of the hierarchy to allow for convergence. The only physical density matrix is the $(0,...,0)$ indexed matrix and refers to the reduced density matrix. All other $\rho^n$ are auxiliary and encapsulate the collective environmental effects. The terms carrying an index $n^\pm_{jk}$ refer to auxiliary density operators with an index raised or lowered by one.

For the case of the interaction Hamiltonian given in Eq.~\ref{eqn:reducedHam}, we have that $Q =\,\sigma_z$. In order to deploy the HEOM framework, we need only characterize the associated spectral density of the environment the biomolecule sits in.

\subsection*{Parameters for GFP system}
The parameters utilized throughout this paper are provided below in Table. \ref{table}.
\
\begin{table}[h]
\centering
\begin{tabular}{|c|c|c|c|} 
 \hline
 \hline
 Parameter & Value (units) \\ 
  \hline
  \hline
 GFP transition wavelength & $490$nm ~\cite{GFPOG}\\ 
 \hline
 $\Delta \mu $ & 10 D ~\cite{Lin20GFP} \\ 

$\mu_T$ & 7.1 D~\cite{Chung_2016TransDip}\\

$\kappa$ & 0.35~\cite{kappavalue}\\

 $r_2$ & $13.35$\AA ~\cite{StructureGFP}\\

 $r_1$ & $5$\AA~\cite{StructureGFP}\\

  $R$ & $27.5$\AA~\cite{Johnson2021-nc}\\

  $n$ & $1.72$~\cite{GFPRefractiveIndex}\\

  $T$ & $293 $K\\

  $\varepsilon_\textrm{p}$ & $4$~\cite{ProteinDielectrics}\\
 \hline
  Single Debye & ~\cite{McKenzieGilmore}\\
 \hline
  $\tau_D$ & $8.2$ps\\

  $\varepsilon_\infty$ & $4.21$\\

  $\varepsilon_\textrm{s}$ & $78.3$\\
 \hline
   Double Debye & ~\cite{DDebye}\\
\hline
  $\tau_D$ & $8.3$ps\\

   $\tau_1$ & $0.36$ps\\

  $\varepsilon_\infty$ & $4.48$\\

  $\varepsilon_1$ & $6.6$\\

  $\varepsilon_\textrm{s}$ & $78.6$\\
 \hline
 \end{tabular}
\caption{Relevant parameters for GFP system. }
\label{table}
\end{table}

\subsection*{Parameters for the Double Debye System}
We introduce here the two  Debye timescale decay model~\cite{DoubleDebyeSolvent} for the solvent media, used to model the  density of states of the electromagnetic field in the high-frequency regime, relevant for transitions in the optical regime. 

We consider the following frequency dependant dielectric~\cite{DDebye}
\begin{equation}
    \varepsilon(\omega) = \varepsilon_\infty +\frac{\varepsilon_\textrm{s} -\varepsilon_1}{1-i\omega \tau_D} + \frac{\varepsilon_1 - \varepsilon_\infty}{1-i\omega \tau_1}, 
\end{equation}
which  leads to spectral densities of the form
\begin{equation}
    J(\omega) = \lambda\frac{a\omega + b\omega^3}{c+d\omega^2+f\omega^4}.
    \label{eqn:SpecDensDDApp}
\end{equation}
The relationships below link the parameters defining the model are related to the electric permittivity (relevant for Model 3)
\begin{align*}
    \lambda =& \frac{\Delta\mu^2}{8\pi^2\varepsilon_0r_1^3}\\
    a =& 27 r_1^3 r_2^3 \varepsilon_\textrm{p}^2 (\varepsilon_{1\infty} \tau_1 + \varepsilon_{s1} \tau_D),\\
    b =& 27 r_1^3 r_2^3 \varepsilon_\textrm{p}^2 \tau_1 \tau_D (\varepsilon_{s1} \tau_1 + \varepsilon_{1\infty} \tau_D),\\
    c =& (2 r_1^3 (-1 + \varepsilon_\textrm{p}) (\varepsilon_{1\infty} + \varepsilon_\infty - \varepsilon_\textrm{p} + \varepsilon_{s1}) + 
  r_2^3 (1 + 2 \varepsilon_\textrm{p}) (2 \varepsilon_{1\infty} + 2 \varepsilon_\infty + \varepsilon_\textrm{p} + 2 \varepsilon_{s1}))^2, \\
   d =& (-4 r_1^3 r_2^3 (-1 - \varepsilon_\textrm{p} + 2 \varepsilon_\textrm{p}^2) (-2 (\varepsilon_{s1} \tau_1 + \varepsilon_{1\infty} \tau_D)^2 - 
     2 \varepsilon_\infty^2 (\tau_1^2 + \tau_D^2) + \varepsilon_\infty \varepsilon_\textrm{p} (\tau_1^2 + \tau_D^2)\nonumber\\& + 
     \varepsilon_\textrm{p}^2 (\tau_1^2 + \tau_D^2) - 4 \varepsilon_\infty (\varepsilon_{s1} \tau_1^2 + \varepsilon_{1\infty} \tau_D^2) + 
     \varepsilon_\textrm{p} (\varepsilon_{s1} \tau_1^2 + \varepsilon_{1\infty} \tau_D^2)) \\&+ 
  r_2^6 (1 + 2 \varepsilon_\textrm{p})^2 (4 (\varepsilon_{s1} \tau_1 + \varepsilon_{1\infty} \tau_D)^2\nonumber\\& + 4 \varepsilon_\infty^2 (\tau_1^2 + \tau_D^2) + 
     4 \varepsilon_\infty \varepsilon_\textrm{p} (\tau_1^2 + \tau_D^2) + \varepsilon_\textrm{p}^2 (\tau_1^2 + \tau_D^2) + 
     8 \varepsilon_\infty (\varepsilon_{s1} \tau_1^2 + \varepsilon_{1\infty} \tau_D^2)\nonumber \\&+ 4 \varepsilon_\textrm{p} (\varepsilon_{s1} \tau_1^2 + \varepsilon_{1\infty} \tau_D^2))+ 
  4 r_1^6 (-1 + \varepsilon_\textrm{p})^2 ((\varepsilon_{s1} \tau_1 + \varepsilon_{1\infty} \tau_D)^2 + \varepsilon_\infty^2 (\tau_1^2 + \tau_D^2)\\& + 
     \varepsilon_\textrm{p}^2 (\tau_1^2 + \tau_D^2) - 2 \varepsilon_\textrm{p} (\varepsilon_{s1} \tau_1^2 + \varepsilon_{1\infty} \tau_D^2) + 
     2 \varepsilon_\infty (\varepsilon_{s1} \tau_1^2 + \varepsilon_{1\infty} \tau_D^2 - \varepsilon_\textrm{p} (\tau_1^2 + \tau_D^2))))\nonumber,\\
     f =& (2 r_1^3 (\varepsilon_\infty - \varepsilon_\textrm{p}) (-1 + \varepsilon_\textrm{p}) + r_2^3 (2 \varepsilon_\infty + \varepsilon_\textrm{p}) (1 + 2 \varepsilon_\textrm{p}))^2 \tau_1^2 \tau_D^2,\\
     \varepsilon_{1\infty} &= \varepsilon_1-\varepsilon_\infty,\\
     \varepsilon_{s1} &= \varepsilon_\textrm{s} -\varepsilon_1. \numberthis
\end{align*}

\subsection*{Mapping HEOM to time-dependant GKSL}
\label{appendix:MapHEOM}
For the single two-level system interacting with the bosonic environment given by 
\begin{equation}
    H = H_A + H_B + H_I,
\end{equation}
where 
\begin{align}
    H_A &= \frac{1}{2}\omega_0 \,\sigma_z,\nonumber\\
    H_B &= \sum_\lambda \omega_\lambda a^\dag_\lambda a_\lambda, \\
    H_I &= \,\sigma_z \sum_\lambda g_\lambda(a^\dag_\lambda + a_\lambda).\nonumber
\end{align}
We note that the populations of the two-level systems free energy eigenstates will be unchanged in any evolution. 
As such we can directly compare any time evolution of the reduced density matrix for the two-level system $\rho_A$ to a GKSL-like equation of the form
\begin{equation}
    \Dot{\rho}_A = -i[H_A(t),\rho_A] + \Gamma \left(\,\sigma_z\rho_A\,\sigma_z -\rho_A \right)
\end{equation}
where the Lindblad rate is time dependant and of the form 
\begin{equation}
    \Gamma(t) = -\frac{1}{2} \frac{\Re{\Dot{\rho}_{12}}}{\Re{\rho_{12}}}.
\end{equation}

For the case of the coupled chromophores, the system Hamiltonian is given by  
\begin{equation}
    H_A  = \Delta \,\sigma_x
\end{equation}
and the write the generic time evolution of the reduced density matrix under the form of a time dependant GKSL equation
\begin{equation}
\Dot{\rho}_A = \sum_{i=\{x,y,z\}} \Gamma_i (\,\sigma_i \rho_A \,\sigma_i - \rho_A),
\end{equation}
where the Lindblad rates 
\begin{align}
\Gamma_z = \frac{1}{4} \left(\frac{2 \Dot{\rho}_{11}}{2 \rho_{11}-1}-\frac{\Im(\Dot{\rho}_{12})}{\Im(\rho_{12})}-\frac{\Re(\Dot{\rho}_{12})}{\Re(\rho_{12})}\right),\\
    \Gamma_x =  \frac{1}{4} \left(\frac{2 \Dot{\rho}_{11}}{1-2 \rho_{11}}-\frac{\Im(\Dot{\rho}_{12})}{\Im(\rho_{12})}+\frac{\Re(\Dot{\rho}_{12})}{\Re(\rho_{12})}\right),\\
    \Gamma_y =\frac{1}{4} \left(\frac{2 \Dot{\rho}_{11}}{1-2 \rho_{11}}+\frac{\Im(\Dot{\rho}_{12})}{\Im(\rho_{12})}-\frac{\Re(\Dot{\rho}_{12})}{\Re(\rho_{12})}\right),
\end{align}
or, equivalently,
\begin{align}
&\Gamma_z = \frac{1}{4}\frac{d}{dt} \left(\ln{\langle \,\sigma_z \rangle}+\ln{\langle \,\sigma_y\rangle}-\ln{\langle \,\sigma_x\rangle}\right),\\
    &\Gamma_x =  \frac{1}{4}\frac{d}{dt} \left(-\ln{\langle \,\sigma_z \rangle}+\ln{\langle \,\sigma_y\rangle}+\ln{\langle \,\sigma_x\rangle}\right),\\
    &\Gamma_y =\frac{1}{4} \frac{d}{dt}\left(-\ln{\langle \,\sigma_z \rangle}-\ln{\langle \,\sigma_y\rangle}-\ln{\langle \,\sigma_x\rangle}\right).
\end{align}
Hence, we can map the dynamics described by the HEOM approach directly onto GKSL-like equations. However, one should note that these Lindblad rates are trajectory dependant due to the explicit dependence on $\rho_A$.

 \printbibliography[heading=subbibliography]
 \end{refsection}

\section*{Data availability}
Source data for all figures are available on
the Figshare data repository (https://doi.org/...). All other data that support  other findings of this study are available from the corresponding author upon reasonable request.

\bigskip

\noindent \textbf{\large Acknowledgments} 

\noindent This work was supported by the Leverhulme Quantum Biology Doctoral Training Centre at the University of Surrey funded by the Leverhulme Trust Training Centre under Grant No. DS-2017–079, and the EPSRC (United Kingdom) Strategic Equipment under Grant No. EP/L02263X/1  (EP/M008576/1) and EPSRC (United Kingdom) under Grant No. EP/M027791/1 awards to M.F. We acknowledge helpful discussions with the members of the Leverhulme Quantum Biology Doctoral Training Centre

\bigskip 

\noindent \textbf{\large Author contributions} 

\noindent A.B. initiated the project,  performed simulations and wrote the paper.  M.F. initiated the programme, oversaw and directed the project and wrote the paper.

\bigskip

\noindent \noindent \textbf{\large Funding} 

\noindent Open access funding provided by the University of Surrey.

\bigskip 

\noindent \textbf{\large Competing interests} 

\noindent The authors declare no competing interests.

\bigskip 

\noindent \textbf{\large Additional information} 

\bigskip 

\noindent  \textbf{Extended data}  is available for this paper at
https://doi.org/....

\bigskip 

\noindent  \textbf{Correspondence and requests for materials}   should be  addressed to Marian Florescu.


 \nocite{*}
 \end{document}